\documentclass[letterpaper,10pt]{article}

\usepackage{titlesec}

\usepackage{color}
\newcommand{\rev}[1]{\textcolor{black}{#1}}

\usepackage{enumerate}
\usepackage{graphicx}
\usepackage{times}
\usepackage{array}
\usepackage[round,numbers]{natbib}
\usepackage{multirow}
\usepackage{booktabs}
\usepackage[small]{caption}
\usepackage[table]{xcolor}

\usepackage{hyperref}
\hypersetup{
    colorlinks,%
    citecolor=black,%
    filecolor=black,%
    linkcolor=black,%
    urlcolor=black
}

\title{\vspace{-0.5in}ppiTrim: constructing non-redundant and up-to-date interactomes}
\author{Aleksandar Stojmirovi\'{c} \ and \ Yi-Kuo Yu$^\ast$\\[.5cm]
\small National Center for Biotechnology Information,\\[-.2cm] \small National Library of Medicine, National Institutes of Health,\\[-.2cm] \small Bethesda, MD 20894, United States\\
 {\small \tt stojmira@ncbi.nlm.nih.gov \textnormal{and} yyu@ncbi.nlm.nih.gov}
}

\begin{document}

\maketitle

\begin{abstract}

\rev{Robust advances in interactome analysis demand comprehensive, non-redundant and consistently annotated datasets. By non-redundant, we mean that the accounting of evidence for every interaction should be faithful: each independent experimental support is counted exactly once, no more, no less. While many interactions are shared among public repositories, none of them contains the complete known interactome for any model organism. In addition, the annotations of the same experimental result by different repositories often disagree. This brings up the issue of which annotation to keep while consolidating evidences that are the same. The iRefIndex database, including interactions from most popular repositories with a standardized protein nomenclature, represents a significant advance in all aspects, especially in comprehensiveness. However, iRefIndex aims to maintain all information/annotation from original sources and requires users to perform additional processing to fully achieve the aforementioned goals. Another issue has to do with protein complexes. Some databases represent experimentally observed complexes as interactions with more than two participants, while others expand them into binary interactions using spoke or matrix model. 
To avoid untested interaction information buildup, it is preferable to replace the expanded protein complexes, either from spoke or matrix models, with a flat list of complex members.
}

\rev{To address these issues and to achieve our goals, we have developed \textit{ppiTrim}, a script that processes iRefIndex to produce non-redundant, consistently annotated datasets of physical interactions. Our script proceeds in three stages: mapping all interactants to gene identifiers and removing all undesired raw interactions, deflating potentially expanded complexes, and reconciling for each interaction the annotation labels among different source databases. As an illustration, we have processed the three largest organismal datasets: yeast, human and fruitfly. 
While \textit{ppiTrim} can resolve most apparent conflicts between different labelings, we also discovered some unresolvable disagreements mostly resulting from different annotation policies among repositories.
} \\
\textbf{\rev{URL:}} {\footnotesize \url{www.ncbi.nlm.nih.gov/CBBresearch/Yu/downloads/ppiTrim.html}}
\end{abstract}

\section*{Introduction}

The current decade has witnessed a significant amount of effort towards discovering the networks of protein-protein interactions (interactomes) in a number of model organisms. These efforts resulted in hundreds of thousands of individual interactions between pairs of proteins being reported~\citep{DF10}. Repositories such as the BioGRID~\citep{SBCB11}, IntAct~\citep{AAAA10}, MINT~\citep{CCLP10}, DIP~\citep{SMSP04}, BIND~\citep{AAAB05,IEB11} and HPRD~\citep{KGKK09} have been established to store and distribute sets of interactions collected from high-throughput scans as well as from curation of individual publications. Depending on its goals, each interaction database, maintained by a different team of curators located around the world includes and annotates interactions differently. Consequently, while many interactions of specific interactomes are shared among databases~\citep{DF10,CYSV09}, no one contains the complete known interactome for any model organism. Constructing a full-coverage protein-protein interaction network therefore requires retrieving and combining entries from many databases.

This task is facilitated by several initiatives developed by the proteomics community over the years. The IMEx consortium~\citep{OKJC07} was formed to facilitate interchange of information between different primary databases by using a standardized format. The Proteomics Standards Initiative Molecular Interaction (PSI-MI) format~\citep{KOMA07} allows a standard way to represent protein interaction information. One of its salient features is the controlled vocabulary of terms that can be used to describe various facets of a protein-protein  interaction including source database, interaction detection method, cellular and experimental roles of interacting proteins and others. The PSI-MI vocabulary is organized as an ontology, a directed acyclic graph (DAG), where nodes correspond to terms and links to relations between terms. This enables the terms to be related in an efficient and algorithm-friendly manner.

Consistently annotated datasets are useful for development and assessment of interaction prediction tools~\citep{MS07,GCW08,KHWC09,LSD10}. Furthermore, such datasets also form the basis of interaction networks, for which numerous analysis tools have been developed~\citep{CTR09,PSS10}. Depending on biological aims of a tool, different entities (nodes) and potentially weighted interactions (edges) may be preferred. The chance of conflicting predictions from different tools can be reduced by starting from a consistently annotated dataset that faithfully represents all available evidences. \rev{Such dataset ought to be comprehensive but also non-redundant: the same experimental evidence for an interaction should appear once and only once.} To maintain a coherent development of biological understanding, it is indispensable to keep the reference datasets up-to-date.

We examined several primary interaction databases with the aim of constructing non-redundant \rev{(in terms of evidence)}, consistently annotated and up-to-date reference datasets of physical interactions for several model organisms.  Unfortunately, the common standard format used by most primary databases still does not allow direct compilation of full non-redundant interactomes. This mainly results from the fact that different primary databases may use different identifiers for interacting proteins and different conventions for representing and annotating each interaction. Combining interaction data from BIND~\citep{AAAB05,IEB11} \rev{(in two versions called `BIND' and `BIND\_Translation')}, BioGRID~\citep{SBCB11}, CORUM~\citep{RWLB10}, DIP~\citep{SMSP04}, HPRD~\citep{KGKK09}, IntAct~\citep{AAAA10}, MINT~\citep{CCLP10}, MPact~\citep{GMOP06}, MPPI~\citep{PKOB05} and OPHID~\citep{BJ05}, the iRefIndex~\citep{RMD08} database represents a significant advance towards a complete and consistent set of all publicly available protein interactions. Apart from being comprehensive and relatively up-to-date, the main contribution of iRefIndex is in addressing the problem of protein identifiers by mapping the sequence of every interactant into a unique identifier that can be used to compare interactants from different source databases. In a further `canonicalization' procedure~\citep{TRTV10}, different isoforms of the same protein are mapped to the same canonical identifier. By adhering to the PSI-MI vocabulary and file format, iRefIndex provides largely standardized annotations for interactants and interactions. Construction of iRefIndex led to the development of iRefWeb, a web interface for interactive access to iRefIndex data~\citep{TRTV10}. iRefWeb allows an easy visualization of evidence for interactions associated with user-selected proteins or publications. Recently, the authors of iRefIndex and iRefWeb published a detailed analysis of agreement between curated interactions within iRefIndex that are shared between major databases~\citep{TRTD10}.

However, aiming to maintain all information from original sources, iRefIndex \rev{requires users to perform additional processing to fully achieve the aforementioned goals. In particular, iRefIndex considers redundancy in terms of (unordered) pairs of interactants rather than in terms of experimental evidence associated with an interaction.} Consequently, there will be features one desires to have that may not fit well within the scope of iRefIndex. For example, one may wish to treat interactions arising from enzymatic reactions as directed and to be able to selectively include/exclude certain types of reactions such as acetylation. In many cases, the information about post-translational modifications is available directly from source databases, but is not integrated into iRefIndex. Another issue that propagates into iRefIndex from source databases has to do with protein complexes. Some databases represent experimentally observed complexes as interactions with more than two participants, while others expand them into binary interactions using spoke or matrix model~\citep{DF10}. \citet{TRTD10} recently observed that this different representation of complexes is responsible for a significant number of disagreements between major databases curating the same publication. From our earlier work~\citep{SY09}, we found that such expanded complexes may lead to nodes with very high degree and often introduce undesirable shortcuts in networks. To fairly treat the information provided by protein complexes without exaggeration, it is preferable to replace the expanded interactions, either from spoke or matrix models, with a flat list of complex members. Additionally, we discovered that the mapping of each protein to a canonical group by iRefIndex would sometimes place protein sequences clearly originating from the same gene (for example differing in one or two amino acids) into different canonical groups.

To achieve the goal of constructing non-redundant, consistently annotated and up-to-date reference datasets, we developed a script, called ppiTrim, that processes iRefIndex and produces a consolidated dataset of physical protein-protein interactions within a single organism.

\section*{Materials and Methods}

Our script, called ppiTrim, is written in the Python programming language. It takes as input a dataset in iRefIndex PSI-MI TAB 2.6 format, with 54 TAB-delimited columns (36 standard and 18 added by iRefIndex). After three major processing steps, it outputs a consolidated dataset, in PSI-MI TAB 2.6 format, containing only the 36 standard columns \rev{(Supplementary Table 1)}. The three processing steps are: (i) mapping all interactants to NCBI Gene IDs and removing all undesired raw interactions; (ii) deflating potentially expanded complexes; and (iii) \rev{collecting all raw interactions, originated from a single publication, that have the same interactants and compatible experimental detection method annotations into one consolidated interaction.} At each step, ppiTrim downloads the files it requires from the public repositories and writes its intermediate results as temporary files.

\subsection*{Phase I: initial filtering and mapping interactants}

\rev{In Phase I, ppiTrim takes the original iRefIndex dataset and classifies each raw interaction (either a binary interaction corresponding to a single line in the input file or a complex supported by several lines) into one of four distinct categories: removed (not examined further), biochemical reaction, complex or potentially part of a complex, and other (direct binary binding interaction). It removes interactions marked as genetic, originating from publications specified through a command line parameter or having interactants from organisms other than the main species of the input dataset (the allowed species can be explicitly provided or any interaction with interactants having different Taxonomy IDs is removed). Additionally, ppiTrim removes all interactions from OPHID and the `original' BIND.} The former is removed because it contains either computationally predicted interactions or interactions verified from the literature using text mining (i.e. without human curation). \rev{The latter is removed because it processes the same original dataset as BIND\_Translation~\citep{IEB11}.}

As a first step, the script seeks to map each interactant to an NCBI Entrez Gene~\citep{MOPT11} identifier. For most interactants, it uses the mapping already provided by iRefIndex. In the cases where iRefIndex provides only a Uniprot~\citep{U10} knowledge base accession, the script attempts to obtain a Gene ID \rev{ in three different ways. First, it searches the iRefIndex \texttt{mappings.txt} file (found compressed in \url{ftp.no.embnet.org/irefindex/data/current/Mappingfiles/}{} for any additional mappings. This part is optional because the \texttt{mappings.txt} file is very large even compressed and it would not be feasible to perform automatic download each time ppiTrim is run. Second, for all unmapped Uniprot IDs, it retrieves the corresponding full Uniprot records using the dbfetch tool from EBI (\url{www.ebi.ac.uk/Tools/dbfetch}{}). If a direct mapping to Gene ID is present within the record as a part of DR field, it is used. Otherwise, the canonical gene name (field GN)} is used to query the NCBI Entrez Gene database for a matching Gene record using an Eutils interface. If a single unambiguous match is found, the record's Gene ID is used for the interactant. \rev{No mapping is performed if multiple matches are obtained.} Every mapped Gene ID is checked against the list of obsolete Gene IDs, which are no longer considered to have a protein product existing \textit{in vivo}. The interactants that cannot be mapped to valid (non-obsolete) Gene IDs are removed along with all raw interactions they participate in.

After assigning Gene IDs, the script considers the PSI-MI ontology terms associated with interaction detection method, interaction type and interactants' biological roles. Using the full PSI-MI ontology file in Open Biomedical Ontology (OBO) format~\citep{SARB07}, it replaces any non-standard terms in these fields (labeled MI:0000) with the corresponding valid PSI-MI ontology terms. The terms marked as obsolete in the PSI-MI OBO file are exchanged for their recommended replacements \rev{(Supplementary Table 2)}. The single exception are the interaction detection method terms for HPRD `in vitro' (MI:0492, translated from MI:0045 label in iRefIndex) and `in vivo' (MI:0493) interactions, which are kept throughout the entire processing.

Source interactions annotated with a descendant of the term MI:0415 (enzymatic study) as their detection method or with a descendant of the term MI:0414 (enzymatic reaction) as their interaction type are classified as candidate biochemical reactions. This category also includes any interactions (including those with more than two interactants) where one of interactants has a biological role of MI:0501 (enzyme) or MI:0502 (enzyme target). In the recent months, the BioGRID database has started to provide additional information about the post-translational modifications associated with the `biochemical activity' interactions, such as phosphorylation, ubiquitination etc. This information is available from the BioGRID datasets in the new TAB2 format but is not yet reflected in the PSI-MI terms for interaction type provided in the PSI-MI 2.5 format or in iRefIndex. Since the post-translational modifications annotated by the BioGRID can be directly matched to standard PSI-MI terms \rev{(Supplementary Table 3)}, the script downloads the most recent BioGRID dataset in TAB2 format, extracts this information and assigns appropriate PSI-MI terms for interaction type to the candidate biochemical reactions from iRefIndex that originate from the BioGRID.

Any source interaction not classified as candidate biochemical reaction is considered for assignment to the candidate complex categories. This category includes all true complexes (having edge type `C' in iRefIndex), interactions having a descendant of MI:0004 (affinity chromatography) as the detection method term or MI:0403 (colocalization) as the interaction type, as well as the interactions corresponding to the BioGRID's `Co-purification' category. Interactions with interaction type MI:0407 (direct interaction) are never considered candidates for complexes. All source interactions not falling into candidate biochemical reaction or candidate complex categories are considered ordinary binary physical interactions.

\subsection*{Phase II: deflating spoke-expanded complexes}

\begin{figure}[h!]
\begin{center}
\includegraphics[scale=0.75]{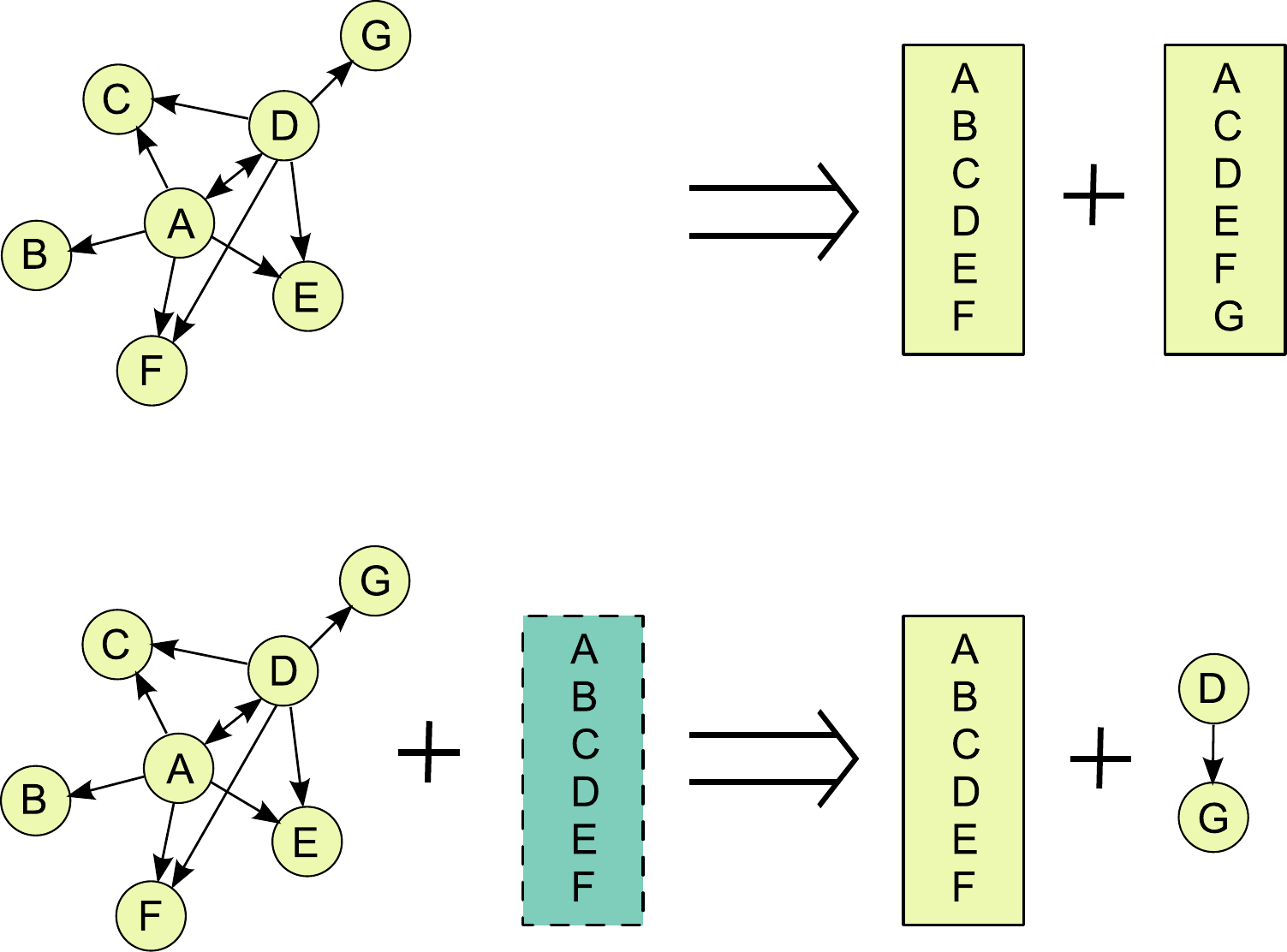}
\caption[Procedures for deflating complexes]{\footnotesize ppiTrim uses two procedures for complex deflation: pattern detection (top) and template matching (bottom). As an example, assume that a graph ABCDEFG, shown on the left, could be constructed from complex candidate interactions annotated by the BioGRID from a single publication. The arrows indicate bait to prey relationships, with the interaction A--D being repeated twice, once with A and once with D as a bait. Pattern detection algorithm (top) would recognize A and D as hubs of potentially spoke-expanded complexes and thus replace all pairwise interactions on the left with complexes ABCDEF and ACDEFG. Suppose that the complex \rev{ACDEF} was reported from the same publication by a different database. Then, template matching procedure (bottom) would generate the complex \rev{ACDEF} (with all other annotation, such as experimental detection method, retained from the original interactions) and remove all original interactions except D--G \rev{and A--B}. After performing both procedures, ppiTrim consolidates the results so that the overall result would be replacing the original interactions by complexes \rev{ACDEF}, ABCDEF and ACDEFG with edge type codes \rev{`R', `A'} and `A', respectively. The \rev{interactions A--B and} D--G would not be retained since \rev{they are} contained within the \rev{deflated complexes ABCDEF and ACDEFG}.
}\label{fig:complexes}
\end{center}
\end{figure}

The Phase II script attempts to detect spoke-expanded complexes from `candidate complex' interactions and deflate them into interactions with multiple interactants. First, all candidate interactions are grouped according to their publication (Pubmed ID), source database, detection method and interaction type. Each group of source interactions is turned into a graph and considered separately for consolidation into one or more complexes. When a portion of a group of interactions is deflated, we replace these source interactions by a complex containing all their participants. \rev{Each collapsed complex is represented using bipartite representation in the output MITAB file (the same as the original complexes from iRefIndex, but using newly generated complex IDs) and the references to the original source interactions are preserved (Supplementary Table 1).} Two procedures are used for consolidation: pattern detection and template matching (Fig.~\ref{fig:complexes}). \rev{The deflation algorithm for each new complex is indicated in the output file through its edge type (Table~\ref{tbl:edgetypes}).}

Pattern detection procedure is used only for the interactions from the BioGRID. Unlike the interactions from the DIP, those interactions are inherently directed since one protein is always labeled as bait and other as prey (in many cases this labeling is unrelated to the actual experimental roles of the proteins). The pattern indicating a possible spoke-expanded complex consists of a single bait being linked to many preys. Since all interactions in the BioGRID's 'Co-purification' and 'Co-fractionation' categories arise from complexes that are spoke-expanded using an arbitrary protein as a bait (BioGRID Administration Team, private communication), a bait linked to two or more preys can in that case always be considered an expanded complex and deflated. Such deflated complexes are assigned the edge type code `G'. The remainder of the complex candidate interactions from the BioGRID were obtained by affinity chromatography and are, in most cases, also derived from complexes. Here we adopted a heuristic that a bait linked to at least three preys can be considered a complex. Clearly, some experiments involve a single bait being used with many independent preys, in which case this procedure would generate a false complex. Therefore, complexes generated in this way are assigned a different edge type code (`A') and the user is able to specify specific publications to be excluded from consideration as well as the maximal size of the complex.

\begin{table}
\rev{
\begin{center}
\caption{\rev{Edge type codes used by ppiTrim \label{tbl:edgetypes}}}
{\footnotesize
\begin{tabular}{lp{13cm}}\toprule
Code & Description \\ \midrule
X & undirected binary interaction (physical binding) \\
D & directed binary interaction (biochemical reaction) \\
B & biochemical reaction without indication of directionality \\
C & original complex (from iRefIndex) \\
G & spoke-expanded complex; deflated by pattern matching from BioGRID's 'Co-purification' and 'Co-fractionation' categories (reliable) \\
R & potential spoke-expanded complex; deflated by template matching of a `C'-complex \\
A & potential spoke-expanded complex (BioGRID only); deflated by pattern detection \\
N & potential spoke-expanded complex; deflated by template matching of a `G'- or `A'-complex \\ \bottomrule
\end{tabular}
}
\end{center}
} % rev
\end{table}

The second procedure is based on matching each group of candidate interactions to the complexes indicated by other databases (templates), mostly from IntAct, MINT, DIP and BIND. In this case, the script checks for each protein in the group whether it, together with all its neighbors, is a superset of a template complex. If so, all the candidate interactions between the proteins within the complex are deflated. The neighborhood graph is undirected for all source databases except the BioGRID. The new complexes generated in this way are given the code `R'. The scripts also attempts to use complexes generated from the BioGRID's interactions through a pattern detection procedure as templates, in which case the newly generated complexes have the code `N'. Any source interactions that cannot be deflated into complexes are retained for Phase III.

\subsection*{Phase III: Normalizing interaction type annotation}

\rev{\subsubsection*{Overview}}

\begin{figure}[h!]
\begin{center}
\includegraphics[scale=0.7]{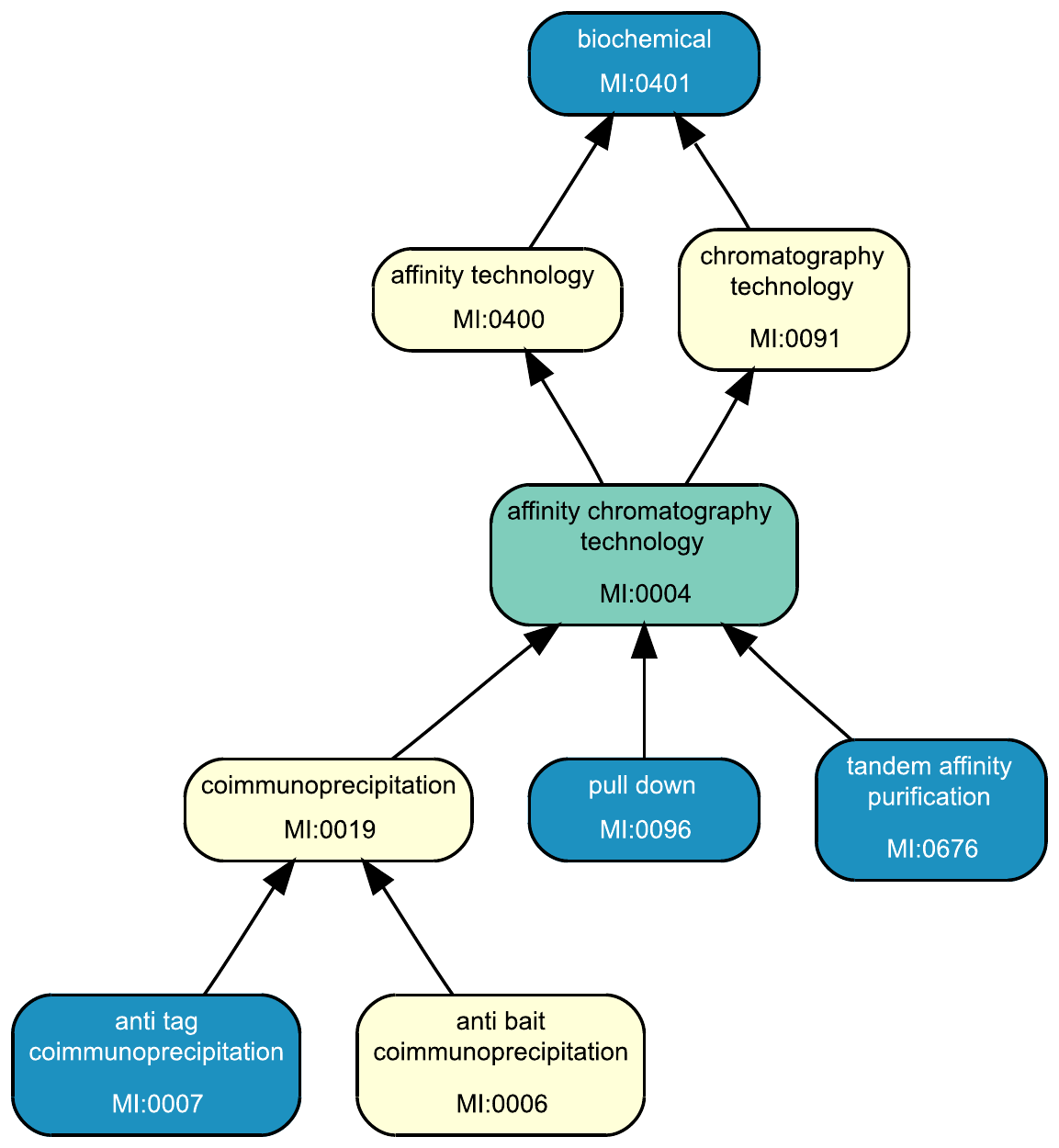}
\caption[Compatible ontology terms.]{\footnotesize The picture shows a part of the PSI-MI ontology graph for interaction detection method associated with a hypothetical cluster of source interactions involving the same interactants from the same publication. The terms colored blue are associated with the source interactions within the cluster, while those marked yellow and green are present in the ontology but do not label any source interaction from the cluster. The entire cluster as shown is consistent, with the term MI:0401 as the maximal element. Its finest consistent term is MI:0004 (colored green) since the cluster members smaller than it are not comparable between themselves. Removing the source interactions labeled by MI:0401 from the cluster would result in three distinct subclusters. If two subclusters contain no interaction from the same source database, they would be reported as conflicts.}\label{fig:example1}
\end{center}
\end{figure}

\rev{The goal of the final phase of ppiTrim is to consolidate all evidence for an interaction, obtained from a single experiment, into one \emph{consolidated interaction} record. Every source publication contains descriptions of one or more experiments that result in reported interactions. Unfortunately, distinct experiments within each publication are not annotated in all source databases, with the exception of the interactions from IntAct and MINT that appear to distinguish experiments using a numbered suffix to the author's name in the `Author' field. It is therefore necessary to rely on the experimental detection method terms to determine whether source records from different databases, with the same interactants and source publication, represent the evidence for the same interaction. Ideally, all such records with the same detection method can be collapsed into one consolidated interaction, although this may undercount multiple evidences from the same publication obtained by distinct experiments. However, different databases have different annotation policies and do not necessarily use the same PSI-MI term to annotate a given experimental method. To resolve detection method term disagreements, we use the PSI-MI ontology structure (Fig.~\ref{fig:example1}). Two compatible terms assigned by different source databases are considered to represent the same experimental method within a publication. These annotated records are thus consolidated.}

\rev{The Phase III algorithm proceeds as follows. All source interactions and complexes (original as well as deflated in Phase II) are divided into `clusters'. 
Interactions that share the same interactants and the source publication are placed into the same cluster. The order of interactants is significant only for biochemical reactions, which are treated as directed interactions (only when direction can be ascertained). Each cluster is processed independently and divided into subclusters based on compatibility of the PSI-MI terms for interaction detection method. Interactions from each subcluster are collected into a single consolidated interaction, which is output to the final dataset. The consolidated record preserves references to all original interactions. Each consolidated interaction is assigned a single PSI-MI term for interaction detection method that most specifically describes the entire collection of annotation terms within the subcluster. For easier reference, each consolidated interaction is given a unique ppiTrim ID, which is similar to RIGID from iRefIndex. This is a SHA1 hash of a dot-separated concatenation of its interactants (Gene IDs), publication(s), detection method, interaction type and edge type. Every complex uses its ppiTrim ID as its primary ID.}

\rev{\subsubsection*{Reconciling annotation}}

The DAG structure of an ontology naturally induces a partial order between the terms: for two terms $u$ and $v$, we say that $u$ refines $v$ ($u$ is smaller $v$, $u$ precedes $v$) if there exists a directed path in the DAG from $u$ to $v$. Two PSI-MI terms can be considered compatible if they are comparable, that is, one refines the other. Every nonempty collection of terms $U$ can be uniquely split into disjoint sets $U_i$, such that every $U_i$ has a single maximal element (an element comparable to and not smaller than any other member) and contains all members of $U$ comparable to its maximal element. Every subcollection $U_i$ is then consistent because there exists at least one term within it that can describe all its members, while any two members from different subcollections are incomparable. The \emph{finest consistent term} of a subcollection $U_i$ is the smallest member of $U_i$ that is comparable to all its members (it can also be defined as the smallest member of the intersection of the transitive closures of all the members of $U_i$.). If $U_i$ is a total order, where all members are pairwise comparable, the finest consistent term is the minimal term. On the other hand, the minimal term need not exist (Fig.~\ref{fig:example1}), so that the finest consistent term is higher in the hierarchy and represents the most specific annotation that can be assigned to $U_i$ as a whole.

To produce consolidated interactions from a single cluster, each of its members (interactions) is identified with its PSI-MI term for information detection method. For every cluster member, the set of all other members with compatible annotations (`compatible set') is computed. As a special case, the following
detection method tags are treated as smaller than any other: `unspecified method' (MI:0686), `in vivo' and `in vitro' (The latter two are from HPRD only).
In this way, non-specific annotations are considered as compatible with all other, more specific evidences.  Compatible sets are further grouped according to their maximal elements. Within each group, the union of the compatible sets produces a subcluster. The finest consistent term for each subcluster is found by considering all PSI-MI terms on the paths from the subcluster members to its maximum -- the search is not restricted to those terms that are within the subcluster (Fig.~\ref{fig:example1}).

\rev{\subsubsection*{Conflicts}}

\rev{We consider two subclusters of the same cluster to be in an unresolvable conflict if there is no source database shared between them.} This definition takes into account that a source database may report an interaction several times for the same publication, using the same or different interaction detection method. If two databases annotate the same interaction using incompatible terms, this is most likely due to an error or specific disagreement about the appropriate label, rather than that each database is reporting a different experiment from the same publication. \rev{Unresolvably conflicting interaction records, after consolidation, point to each other using ppiTrim ID in the `Confidence' field.}

\rev{ppiTrim also collects statistics about resolvable conflicts in its temporary output files. A resolvable conflict is the case where source interactions within a single subcluster have compatible but different experimental detection method labels.}

\subsection*{Evaluation of the script}

To test ppiTrim, we applied it to the yeast (\textit{S. cerevisiae}), human (\textit{H. sapiens}) and fruitfly (\textit{D. melanogaster}) datasets from iRefIndex release 8.0-beta, dated Jan 19th 2011. \rev{The script was run on June 13th 2011 and used the then-current versions of Uniprot and NCBI Gene databases. We restricted protein interactors to allowed NCBI Taxonomy IDs: 4932 and 559292 for yeast, 9606 for human, and 7227 for fruitfly datasets.} When processing the yeast dataset, we accounted for two special cases. First, we specifically removed the genetic interactions reported by \citet{TLBD04} because they were not labeled as genetic for all source databases. Second, we excluded the dataset by \citet{CKZG07} from Phase II and retained all its interactions as binary undirected. \rev{This dataset is present only in the BioGRID and can be considered computationally derived and partially redundant. \citet{CKZG07} reprocessed the data from \citet{GAGK06} and \citet{KCYZ06} to obtain an improved set of pairwise interactions. \citet{CKZG07} used hierarchical clustering to recover protein complexes, but these are not present in the BioGRID. In spite of its redundancy, we decided not to entirely remove this dataset but also not to attempt to deflate its potential complexes because bait/prey assignments may not be meaningful in this case.}

\section*{Results and Discussion}

The results of applying ppiTrim to process iRefIndex 8.0 are shown in Tables \ref{tbl:mapping2} -- \ref{tbl:consolidate}. \rev{The statistics of ID mapping (Tables \ref{tbl:mapping2} and \ref{tbl:mapping1}) show that a considerable number of interactants could be additionally mapped to Gene ID in human and fruitfly datasets, thus enabling us to take into consideration a few thousand of raw interactions that would otherwise be filtered. This is also evident in terms of iRefIndex RIGIDs (Supplementary Table 4), which associate all raw interactions with interactants with same sequences to a single record. For yeast, the number of interactions gained by mapping to Gene IDs is small because most of mapped IDs were not valid.}

\begin{table}
\begin{center}
\caption{Processing source interactions \label{tbl:mapping2}}
\rev{
{\footnotesize
\begin{tabular}{lrrrrr} \toprule
Species & Initial & Removed & Without Gene ID & Retained & With Mapped Gene ID \\ \midrule
\textit{S. cerevisiae} & 400449 & 173815 & 3608 & 223026 & 880 \\
\textit{H. sapiens} & 382094 & 148724 & 2738 & 230632 & 16187 \\ 
\textit{D. melanogaster}  & 154770 & 32477 & 9476 & 112817 & 3427 \\  \bottomrule
\end{tabular}
}
} % rev
\caption*{\footnotesize\rev{Statistics of initial processing of raw interactions from iRefIndex. Shown are the initial number, total number removed due to filtering criteria, number removed due to missing Gene ID, total number of retained and the number retained containing at least one interactant with mapped Gene ID.}}
\end{center}
\end{table}

\begin{table}
\begin{center}
\caption{Mapping CROGID identifiers from iRefIndex into Gene IDs \label{tbl:mapping1}}
\rev{
{\footnotesize
\begin{tabular}{l>{\raggedleft}m{1.2cm}>{\raggedleft}m{1.2cm}>{\raggedleft}m{1.2cm}>{\raggedleft}m{1.2cm}>{\raggedleft}m{1.2cm} %
>{\raggedleft}m{1.2cm} >{\raggedleft}m{1.4cm}} \toprule
\multirow{2}{*}{Species} & \multicolumn{3}{c}{Initial CROGIDs} & \multicolumn{2}{c}{Aditional Mapped} & \multicolumn{2}{c}{Final} %
 \tabularnewline 
& total & mapped & orphans & total & valid & CROGIDs & Gene IDs \tabularnewline \midrule
\textit{S. cerevisiae} & 6159 & 5552 & 607 & 433 & 47 & 5599 & 5618 \tabularnewline
\textit{H. sapiens} & 14047 & 11432 & 2615 & 1261 & 1261 & 12693 & 11786 \tabularnewline
\textit{D. melanogaster} & 9379 & 7810 & 1569 & 566 & 566 & 8346 & 7846 \tabularnewline \bottomrule
\end{tabular}
}
} % rev
\caption*{\footnotesize \rev{Statistics of mapping CROGIDs into Gene IDs. Columns 2-4 show the total number of CROGIDs considered, the number that could be directly mapped to GeneIDs and the number of `orphans' that are not associated with a Gene ID in the iRefIndex file. Columns 5 and 6 show the numbers of CROGIDs additionally mapped to GeneIDs, while the last two columns show the final number of CROGIDs accepted and the corresponding number of Gene IDs. It is possible for a CROGID to map to multiple Gene IDs (if multiple genes encode the same protein sequence) as well as for multiple CROGIDs to map to a single GeneID (if our additional mapping links them to the same gene).}}
\end{center}
\end{table}

We chose to standardize proteins using NCBI Gene identifiers rather than the iRefIndex-provided canonical IDs (CROGIDs) for several reasons. NCBI Gene records not only associate each gene with a set of reference sequences, but also include a wealth of additional data (e.g. list of synonyms) and links to other databases such as Gene Ontology~\citep{ABB00} that are important when using the interaction dataset in practice. In addition, Gene records are regularly updated and their status evaluated based on new evidence. Thus, a gene record may be split into several new records or marked as obsolete if it corresponds to an ORF that is known not to produce a protein. For network analysis applications, it is desirable that only the proteins actually expressed in the cell are represented in the network and hence the gene status provided by NCBI Gene is a valuable filtering criterion. \rev{Our results in yeast (Table~\ref{tbl:mapping1}) support this premise: most CROGIDs without Gene ID are associated with sequences derived from ORFs that were subsequently declassified as genes.} However, CROGIDs do have one advantage over NCBI Gene IDs in that they are protein-based and hence identical protein products of several genes (like histones) are clustered together.

\rev{There are several reasons that our algorithm was able to introduce many additional associations of CROGIDs to Gene IDs. First, iRefIndex only provides mappings to Gene IDs for interactors that have a sequence that exactly matches a sequence in an NCBI RefSeq record (Ian Donaldson, private communication). By a case-by-case examination of some orphaned yeast sequences that could be mapped to Gene ID, we found that they were orphans because they differed in one or two amino acids from that protein's reference representative in RefSeq but were not clustered with that representative's Gene record. Additional mappings can be found through database cross-reference from a Uniprot record pointing to a Gene ID. The iRefIndex canonicalization procedure captures some of these associations in the \texttt{mappings.txt} file but they are not available in the main iRefIndex MITAB files. We have found (Supplementary Table 5) that some CROGIDs (mostly in human) can be additionally mapped by using this information in the \texttt{mappings.txt} file. Notably, ppiTrim accesses a more recent version of Uniprot then iRefIndex and is thus able to find more mappings by accessing Uniprot cross-references directly. Finally, there is a substantial number of Uniprot records that do not have a cross-reference to NCBI Gene but can be linked to a Gene record through their canonical gene names.  This last approach can be suggested as an improvement for iRefIndex canonicalization processing.
}

\rev{Around 10\% of CROGIDs could not be mapped to Gene IDs even after processing with ppiTrim algorithms. A few interactors (Supplementary Table 5) have only PDB accessions as their primary IDs since their interactions were derived from crystal structures. In such cases, often only partial sequences of participating proteins are available. These partial sequences cannot be fully matched to any Uniprot or RefSeq record and hence are assigned a separate ID. Hence, an improvement for our procedure, that would account for this case as well as for those unmapped proteins that differ from canonical sequences only by few amino acids, would be to use direct sequence comparison to find the closest valid reference sequence. This task may not be technically difficult (a similar procedure was applied by \citet{AOY08} to construct protein databases for mass spectrometry data analysis) but is beyond the scope of ppiTrim, which is intended as a relatively short standalone script. In our opinion, such additional mappings would best be performed at the level of reference sequence databases such as Uniprot or RefSeq, which contain curator expertise to resolve ambiguous cases.}

\begin{table}
\begin{center}
\caption{Deflating spoke-expanded complexes \label{tbl:collapsing}}
\rev{
{\footnotesize
\begin{tabular}{lrrr>{\raggedleft}p{0.7cm}>{\raggedleft}p{0.7cm}>{\raggedleft}p{0.7cm}>{\raggedleft}p{0.7cm}>{\raggedleft}p{0.7cm}}\toprule
\multirow{2}{*}{Species} & \multirow{2}{*}{Publications} & \multicolumn{2}{c}{Pairs} & \multicolumn{5}{c}{Complexes} \tabularnewline
& & initial & remaining & \centering C & \centering G & \centering R & \centering A & \centering N \tabularnewline \midrule
\textit{S. cerevisiae} & 3924 & 118819 & 28643 & 7729 & 323 & 5384 & 3190 & 1311 \tabularnewline
\textit{H. sapiens} & 10317 & 56111 & 35650 & 8382 & 181 & 1143 & 1443 & 304 \tabularnewline
\textit{D. melanogaster} & 398 & 1722 & 1053 & 220 & 16 & 82 & 33 & 3 \tabularnewline \bottomrule
\end{tabular}
}
} % rev
\caption*{\footnotesize Shown are the numbers of complexes obtained by deflating binary interactions with affinity chromatography (or related) as experimental method. Types of complexes are indicated by one letter codes described in \rev{Table~\ref{tbl:edgetypes}}. The counts of pairs shown include those from publications with fewer than three interactions (per database), which could never be deflated into complexes.}
\end{center}
\end{table}

Protein complexes obtained through chromatography techniques provide information complementary to direct binary interactions. While it is often difficult to determine the exact layout of within-complex pairwise interactions, an identification of an association of several proteins using mass spectroscopy is an evidence for \textit{in vivo} existence of that association. Unfortunately, in spite of its great importance, the currently available information within iRefIndex is deficient because of different treatments of complexes by different source databases. Our results (Table~\ref{tbl:collapsing}) show that the apparently inflated complexity of interaction datasets can be substantially reduced by attempting to collapse spoke-expanded complexes. \rev{For yeast, this results in almost three quarters reduction of the number of candidate interactions. The majority of new complexes falls into `G' and `R' categories, which can be considered most reliable. For the human dataset, reduction is small as a proportion although in absolute terms the number of new complexes is over 3000. The fruitfly dataset did not contain many candidate interactions or complexes and hence not many new complexes were obtained. }

\rev{In general, it is difficult to assess whether newly generated complexes from `A' and `N' categories are biologically justified, that is, whether they represent a functional entity. If a bait and its preys genuinely originate from a single experiment, they definitely form a physical association that may be a part of or an entire functional complex. Since ppiTrim preserves the experimental role labels and the original interaction identifiers, little information is lost by deflating such associations into a single record. On the other hand, for some publications, especially those involving experiments with ubiquitin-like proteins as bait, each bait-prey association may represent a separate experiment and it does not substantiate that different prey proteins may be co-present in the cell. For example, BioGRID provides 158 physical associations from the paper by \citet{HLKL05}, each involving the yeast Smt3p (SUMO, a ubiquitin-like) protein as a bait. In this case, it is not true that all the involved preys together form a large complex with the bait. ppiTrim avoids this particular case by not deflating potentially too large complexes (the maximum deflated complex size is tunable by the user with the default of 120 proteins), but one can assume that some of deflated `complexes' do not exist \textit{in vivo}.}

\rev{To more closely investigate the fidelity of generated complexes, we randomly sampled 25 `A' and `N' deflated yeast complexes from the final output of ppiTrim and examined their original publications. Out of these 25 complexes, 15 originated from high-throughput publications (mostly \citet{GAGK06} and \citet{KCYZ06} -- Supplementary Table 6), while 10 came from small experiments (Supplementary Table 7). In all high-throughput cases, the deflated complex represents a true experimental association. In the cases when authors present their own derived complexes, which in many cases can be found separately under the `C' category, our deflated complexes form parts of larger derived complexes. Indeed, such derived complexes are obtained by assembling the results of several bait-prey experiments, each of which forms a single deflated complex. The results are more varied for low-throughput publications. In most cases, deflated complexes clearly correspond to functional complexes, although it is sometimes difficult to fully relate author's conclusions with their reported results. In two cases, the inferred association is incorrect due to curation errors in the original database. We have also found a single case where the publication authors directly state that proteins in a deflated complex do not form a stable complex.}

\rev{While our sample is extremely small, it does indicate several issues arising from deflation of bait-prey relationships. In most cases, deflated complexes form parts of what are believed to be functional complexes. It appears that curation errors or ambiguities may be a more significant source of wrongly inferred associations than our main assumption that a bait with several preys in a single publication represents a single unit. Overall,} we feel that the benefits from  reduction of interactome complexity outweigh the disadvantages from potentially over deflating interactions. The best way to solve \rev{the problem of different representations of protein complexes} would be at the level of source databases (BioGRID in particular), by reexamining the original publications. Our complexes from the `R' category, where deflated complexes fully agree with an annotated complex from a different database, could serve as a guide in this case.

\begin{table}
\begin{center}
\caption{Final consolidated datasets\label{tbl:consolidate}}
\rev{
{\footnotesize
\begin{tabular}{lrrrrrrrr} \toprule
\multirow{2}{*}{Species} & \multirow{2}{*}{Publications} & \multicolumn{2}{c}{Input Pairs} & \multicolumn{3}{c}{Consolidated} & \multicolumn{2}{c}{Conflicts} \\
& & biochem & other & complexes & directed & undirected & resolvable & unresolvable \\ \midrule
\textit{S. cerevisiae} & 6303 & 5780 & 119329 & 10778 & 5525 & 63648 & 19344 & 454 \\
\textit{H. sapiens} & 22660 & 2446 & 199094 & 6483 & 2042 & 85480 & 26478 & 1333 \\
\textit{D. melanogaster} & 564 & 51 & 111862 & 227 & 33 & 27981 & 19430 & 11 \\ \bottomrule
\end{tabular}
}
} % rev
\caption*{\footnotesize  For each species, shown are the numbers of input pairs (input complexes are those from Table~\ref{tbl:collapsing}), classified as either biochemical reactions (potentially directed) or others; also shown are the final numbers of consolidated interactions (classified as complexes, directed or undirected). The `other' column accounts only for those interactions that were not deflated into complexes in Phase II. \rev{The last two columns show the total numbers of resolvable and unresolvable conflicts between consolidated interactions.} \rev{An unresolvable} conflict is an instance where two consolidated interactions, originated from the same publication, are reported using incompatible experimental detection method labels by different databases. \rev{A resolvable conflict is the case where source interactions within a single consolidated interaction have different (but compatible) experimental detection method labels.}}
\end{center}
\end{table}

\begin{table}
\begin{center}
\caption{Most common interaction detection method PSI-MI term conflicts \label{tbl:conflicts}}
\rev{
{\scriptsize
\begin{tabular}{p{4.8cm}p{1.2cm}p{5.1cm}p{1.5cm}r} \toprule
Term A & Sources A & Term B & Sources B & Counts \\ \midrule
MI:0007 (anti tag coimmunoprecipitation) & M & MI:0676 (tandem affinity purification) & DI & 132 \\
MI:0004 (affinity chromatography) & B & MI:0363 (inferred by author) & I & 60 \\
MI:0018 (two hybrid) & DIMN & MI:0096 (pull down) & BI & 43 \\
MI:0071 (molecular sieving) & DIN & MI:0096 (pull down) & B & 32 \\
MI:0030 (cross-linking study) & DIMN & MI:0096 (pull down) & B & 22 \\
\midrule
MI:0007 (anti tag coimmunoprecipitation) & IM & MI:0676 (tandem affinity purification) & DI & 1227 \\ 
MI:0018 (two hybrid) & BDHIM & MI:0096 (pull down) & BM & 17 \\ 
MI:0096 (pull down) & B & MI:0107 (surface plasmon resonance) & DM & 6 \\ 
MI:0008 (array technology) & I & MI:0049 (filter binding) & M & 5 \\ 
MI:0019 (coimmunoprecipitation) & IM & MI:0096 (pull down) & BI & 5 \\ 
\bottomrule
\end{tabular}
}
} % rev
\caption*{\footnotesize Top five most common interaction detection method PSI-MI term unresolvable conflicts for yeast (top) and human (bottom) datasets are shown. Source databases are indicated by one letter codes B (BioGRID), D (DIP), I (IntAct), H (HPRD), M (MINT),  P (MPPI).}
\end{center}
\end{table}

Overall, our processing significantly reduced the number of interactions within each of the three datasets considered (Table~\ref{tbl:consolidate}). This indicates a significant redundancy, particularly for protein complexes, original and deflated (compare Table~\ref{tbl:collapsing} with Table~\ref{tbl:consolidate}), and for binary interactions. The directed interactions (biochemical reactions) are relatively rarer and largely non-redundant at this stage. Given their importance in elucidating biological function, the directed interactions are expected to be discovered more fully with time. \rev{However, one should note that PSI-MI format can only represent a static relationship among a set of physical entities involved in the same event, but cannot actually represent two sides of a reaction e.g. $A+B \to C + D$. Certain pairs of PSI-MI biological role terms can be combined to represent interaction direction e.g. ‘enzyme’ and ‘enzyme target’, but these are weak compared to the rich ways that pathway databases like Reactome~\cite{COWH11} represent events.}

\rev{To demonstrate the utility of our conflict resolution method, we present the counts for resolvable and unresolvable conflicts in Table~\ref{tbl:consolidate}. Resolvable conflicts significantly outnumber the unresolvable ones. Examining the most common examples of resolvable conflicts (Supplementary Table 8), one can see that a majority of them indeed represent the same experiment. Possible exceptions are human interactions annotated by HPRD, which have ambiguous detection method labels. To address this and similar problems, ppiTrim provides the \texttt{maxsources} confidence score (Supplementary Table 1), which is an estimate of the maximal number of independent experiments contributing to a consolidated interaction. An interesting example of a resolvable conflict in Supplementary Table 8 is the 444 instances of a consolidated interaction containing source interactions with detection method labels MI:0004 (affinity chromatography technology), MI:0007 (anti tag coimmunoprecipitation), and MI:0676 (tandem affinity purification). This case is very similar to the one described in Figure~\ref{fig:example1}: the last two terms are incompatible but the first resolves the conflict as the finest consistent term. }

Upon closer examination \rev{of the few unresolvable conflicts} (Table~\ref{tbl:conflicts}), it can be seen that most common conflicts arise as instances of few specific labeling disagreements between databases. In many cases, such disagreements arise from using different sub-terms of affinity chromatography (see Fig.~\ref{fig:example1}) and can be resolved by assigning a more general term consistent with both conflicting terms. In many other cases, the conflicts are due to BioGRID internally using a more restricted detection method vocabulary than the IMEx databases (DIP, IntAct and MINT). \rev{However, in some rare cases, an unresolvable conflict arises when different databases annotate different experiments from the same publication. For example, each of DIP, BioGRID and IntAct report several raw interactions from the paper by \citet{BT99} (pubmed:9799240), where yeast Met4p protein interacts with each of Met28p, Met31p and Met32p in binary interactions. The paper reports several experiments using different techniques including northern blotting, yeast two hybrid and electrophoretic mobility shift assays. For the interaction between Met4p and Met28p, BioGRID and IntAct report only MI:0018 (yeast two hybrid) method, while DIP reports only MI:0404 (comigration in non denaturing gel electrophoresis), resulting in unresolvable conflict. Hence, in this case, each database on its own provides incomplete evidence for this interaction.}

The ppiTrim algorithms work best if accurate and fully populated fields for interaction detection method, publication and interaction type are available in its input dataset. This requirement is mostly fulfilled. Nevertheless, we have noticed two minor inconsistencies. The first, which will be fixed in a subsequent release of iRefIndex (Ian Donaldson, private communication), involves the PSI-MI labels for interaction detection method for CORUM interactions and complexes. These are missing from iRefIndex although they are present in the original CORUM source files. The second issue concerns missing or invalid Pubmed IDs for certain interactions. We found that a number of interactions with missing Pubmed IDs come from MINT. Upon inspection of the original MINT files, we discovered that in many cases MINT supplies a Digital Object Identifier (DOI) for a publication as its identifier instead of a Pubmed ID (although the corresponding Pubmed ID can be obtained from the MINT web interface). To ensure consistency with other source databases within iRefIndex, it would be desirable to have the Pubmed IDs available for these interactions as well.

In this paper, we have identified the tasks needed for using combined interaction datasets provided by iRefIndex as a basis for construction of reference networks and developed a script to process them into consistent consolidated datasets. We see ppiTrim as answering a \rev{temporary} need for a consolidated database and hope that most of the issues that required processing will be eventually fixed in upstream databases and distributed through IMEx consortium.  At this stage we have not addressed the issue of quality of interactions although such information is available in some databases for some publications~\citep{TRTV10}. Utilizing the quality information in consolidating datasets demands a universal data-quality measure that is not yet existent.

%% ********************* BIBTEX BIBLIOGRAPHY **************************
%\bibliographystyle{dbnatbib}
%\bibliography{ppidb}

\begin{thebibliography}{37}
\providecommand{\natexlab}[1]{#1}
\providecommand{\url}[1]{\texttt{#1}}
\expandafter\ifx\csname urlstyle\endcsname\relax
  \providecommand{\doi}[1]{doi: #1}\else
  \providecommand{\doi}{doi: \begingroup \urlstyle{rm}\Url}\fi

\bibitem[De~Las~Rivas and Fontanillo(2010)]{DF10}
De~Las~Rivas, J. and Fontanillo, C.
\newblock Protein-protein interactions essentials: key concepts to building and
  analyzing interactome networks.
\newblock \emph{PLoS Comput Biol}, 6\penalty0 (6):\penalty0 e1000807, 2010.

\bibitem[Stark\emph{ et~al.}(2011)Stark, Breitkreutz, Chatr-Aryamontri,
  Boucher, Oughtred, Livstone, Nixon, Van~Auken, Wang, Shi, Reguly, Rust,
  Winter, Dolinski, and Tyers]{SBCB11}
Stark, C., Breitkreutz, B.-J., Chatr-Aryamontri, A. \emph{et~al.}
\newblock The {BioGRID} interaction database: 2011 update.
\newblock \emph{Nucleic Acids Res}, 39\penalty0 (Database issue):\penalty0
  D698--704, 2011.

\bibitem[Aranda\emph{ et~al.}(2010)Aranda, Achuthan, Alam-Faruque, Armean,
  Bridge, Derow, Feuermann, Ghanbarian, Kerrien, Khadake, Kerssemakers, Leroy,
  Menden, Michaut, Montecchi-Palazzi, Neuhauser, Orchard, Perreau, Roechert,
  van Eijk, and Hermjakob]{AAAA10}
Aranda, B., Achuthan, P., Alam-Faruque, Y. \emph{et~al.}
\newblock The {IntAct} molecular interaction database in 2010.
\newblock \emph{Nucleic Acids Res}, 38\penalty0 (Database issue):\penalty0
  D525--31, 2010.

\bibitem[Ceol\emph{ et~al.}(2010)Ceol, Chatr-Aryamontri, Licata, Peluso,
  Briganti, Perfetto, Castagnoli, and Cesareni]{CCLP10}
Ceol, A., Chatr-Aryamontri, A., Licata, L. \emph{et~al.}
\newblock {MINT}, the molecular interaction database: 2009 update.
\newblock \emph{Nucleic Acids Res}, 38\penalty0 (Database issue):\penalty0
  D532--9, 2010.

\bibitem[Salwinski\emph{ et~al.}(2004)Salwinski, Miller, Smith, Pettit, Bowie,
  and Eisenberg]{SMSP04}
Salwinski, L., Miller, C.~S., Smith, A.~J. \emph{et~al.}
\newblock The database of interacting proteins: 2004 update.
\newblock \emph{Nucleic Acids Res}, 32\penalty0 (Database issue):\penalty0
  D449--51, 2004.

\bibitem[Alfarano\emph{ et~al.}(2005)Alfarano, Andrade, Anthony, Bahroos,
  Bajec, Bantoft, Betel, Bobechko, Boutilier, Burgess, Buzadzija, Cavero,
  D'Abreo, Donaldson, Dorairajoo, Dumontier, Dumontier, Earles, Farrall,
  Feldman, Garderman, Gong, Gonzaga, Grytsan, Gryz, Gu, Haldorsen, Halupa, Haw,
  Hrvojic, Hurrell, Isserlin, Jack, Juma, Khan, Kon, Konopinsky, Le, Lee, Ling,
  Magidin, Moniakis, Montojo, Moore, Muskat, Ng, Paraiso, Parker, Pintilie,
  Pirone, Salama, Sgro, Shan, Shu, Siew, Skinner, Snyder, Stasiuk, Strumpf,
  Tuekam, Tao, Wang, White, Willis, Wolting, Wong, Wrong, Xin, Yao, Yates,
  Zhang, Zheng, Pawson, Ouellette, and Hogue]{AAAB05}
Alfarano, C., Andrade, C.~E., Anthony, K. \emph{et~al.}
\newblock The {Biomolecular Interaction Network Database} and related tools
  2005 update.
\newblock \emph{Nucleic Acids Res}, 33\penalty0 (Database issue):\penalty0
  D418--24, 2005.

\bibitem[Isserlin\emph{ et~al.}(2011)Isserlin, El-Badrawi, and Bader]{IEB11}
Isserlin, R., El-Badrawi, R.~A., and Bader, G.~D.
\newblock The {Biomolecular Interaction Network Database} in {PSI-MI} 2.5.
\newblock \emph{Database (Oxford)}, 2011:\penalty0 baq037, 2011.

\bibitem[Keshava~Prasad\emph{ et~al.}(2009)Keshava~Prasad, Goel, Kandasamy,
  Keerthikumar, Kumar, Mathivanan, Telikicherla, Raju, Shafreen, Venugopal,
  Balakrishnan, Marimuthu, Banerjee, Somanathan, Sebastian, Rani, Ray,
  Harrys~Kishore, Kanth, Ahmed, Kashyap, Mohmood, Ramachandra, Krishna,
  Rahiman, Mohan, Ranganathan, Ramabadran, Chaerkady, and Pandey]{KGKK09}
Keshava~Prasad, T.~S., Goel, R., Kandasamy, K. \emph{et~al.}
\newblock {Human Protein Reference Database} -- 2009 update.
\newblock \emph{Nucleic Acids Res}, 37\penalty0 (Database issue):\penalty0
  D767--72, 2009.

\bibitem[Cusick\emph{ et~al.}(2009)Cusick, Yu, Smolyar, Venkatesan, Carvunis,
  Simonis, Rual, Borick, Braun, Dreze, Vandenhaute, Galli, Yazaki, Hill, Ecker,
  Roth, and Vidal]{CYSV09}
Cusick, M.~E., Yu, H., Smolyar, A. \emph{et~al.}
\newblock Literature-curated protein interaction datasets.
\newblock \emph{Nat Methods}, 6\penalty0 (1):\penalty0 39--46, 2009.

\bibitem[Orchard\emph{ et~al.}(2007)Orchard, Kerrien, Jones, Ceol,
  Chatr-Aryamontri, Salwinski, Nerothin, and Hermjakob]{OKJC07}
Orchard, S., Kerrien, S., Jones, P. \emph{et~al.}
\newblock Submit your interaction data the {IMEx} way: a step by step guide to
  trouble-free deposition.
\newblock \emph{Proteomics}, 7 Suppl 1:\penalty0 28--34, 2007.

\bibitem[Kerrien\emph{ et~al.}(2007)Kerrien, Orchard, Montecchi-Palazzi,
  Aranda, Quinn, Vinod, Bader, Xenarios, Wojcik, Sherman, Tyers, Salama, Moore,
  Ceol, Chatr-Aryamontri, Oesterheld, Stümpflen, Salwinski, Nerothin, Cerami,
  Cusick, Vidal, Gilson, Armstrong, Woollard, Hogue, Eisenberg, Cesareni,
  Apweiler, and Hermjakob]{KOMA07}
Kerrien, S., Orchard, S., Montecchi-Palazzi, L. \emph{et~al.}
\newblock Broadening the horizon--level 2.5 of the {HUPO-PSI} format for
  molecular interactions.
\newblock \emph{BMC Biol}, 5:\penalty0 44, 2007.

\bibitem[Markowetz and Spang(2007)]{MS07}
Markowetz, F. and Spang, R.
\newblock Inferring cellular networks--a review.
\newblock \emph{BMC Bioinformatics}, 8 Suppl 6:\penalty0 S5, 2007.

\bibitem[Gomez\emph{ et~al.}(2008)Gomez, Choi, and Wu]{GCW08}
Gomez, S.~M., Choi, K., and Wu, Y.
\newblock Prediction of protein-protein interaction networks.
\newblock \emph{Curr Protoc Bioinformatics}, Chapter 8:\penalty0 Unit 8.2,
  2008.

\bibitem[Kanaan\emph{ et~al.}(2009)Kanaan, Huang, Wuchty, Chen, and
  Izaguirre]{KHWC09}
Kanaan, S.~P., Huang, C., Wuchty, S. \emph{et~al.}
\newblock Inferring protein-protein interactions from multiple protein domain
  combinations.
\newblock \emph{Methods Mol Biol}, 541:\penalty0 43--59, 2009.

\bibitem[Lewis\emph{ et~al.}(2010)Lewis, Saeed, and Deane]{LSD10}
Lewis, A. C.~F., Saeed, R., and Deane, C.~M.
\newblock Predicting protein-protein interactions in the context of protein
  evolution.
\newblock \emph{Mol Biosyst}, 6\penalty0 (1):\penalty0 55--64, 2010.

\bibitem[Chautard\emph{ et~al.}(2009)Chautard, Thierry-Mieg, and
  Ricard-Blum]{CTR09}
Chautard, E., Thierry-Mieg, N., and Ricard-Blum, S.
\newblock Interaction networks: from protein functions to drug discovery. a
  review.
\newblock \emph{Pathol Biol (Paris)}, 57\penalty0 (4):\penalty0 324--33, 2009.

\bibitem[Przytycka\emph{ et~al.}(2010)Przytycka, Singh, and Slonim]{PSS10}
Przytycka, T.~M., Singh, M., and Slonim, D.~K.
\newblock Toward the dynamic interactome: it's about time.
\newblock \emph{Brief Bioinform}, 11\penalty0 (1):\penalty0 15--29, 2010.

\bibitem[Ruepp\emph{ et~al.}(2010)Ruepp, Waegele, Lechner, Brauner,
  Dunger-Kaltenbach, Fobo, Frishman, Montrone, and Mewes]{RWLB10}
Ruepp, A., Waegele, B., Lechner, M. \emph{et~al.}
\newblock {CORUM}: the comprehensive resource of mammalian protein
  complexes--2009.
\newblock \emph{Nucleic Acids Res}, 38\penalty0 (Database issue):\penalty0
  D497--501, 2010.

\bibitem[Güldener\emph{ et~al.}(2006)Güldener, Münsterkötter, Oesterheld,
  Pagel, Ruepp, Mewes, and Stümpflen]{GMOP06}
Güldener, U., Münsterkötter, M., Oesterheld, M. \emph{et~al.}
\newblock {MPact}: the {MIPS} protein interaction resource on yeast.
\newblock \emph{Nucleic Acids Res}, 34\penalty0 (Database issue):\penalty0
  D436--41, 2006.

\bibitem[Pagel\emph{ et~al.}(2005)Pagel, Kovac, Oesterheld, Brauner,
  Dunger-Kaltenbach, Frishman, Montrone, Mark, Stümpflen, Mewes, Ruepp, and
  Frishman]{PKOB05}
Pagel, P., Kovac, S., Oesterheld, M. \emph{et~al.}
\newblock The {MIPS} mammalian protein-protein interaction database.
\newblock \emph{Bioinformatics}, 21\penalty0 (6):\penalty0 832--4, 2005.

\bibitem[Brown and Jurisica(2005)]{BJ05}
Brown, K.~R. and Jurisica, I.
\newblock Online predicted human interaction database.
\newblock \emph{Bioinformatics}, 21\penalty0 (9):\penalty0 2076--82, 2005.

\bibitem[Razick\emph{ et~al.}(2008)Razick, Magklaras, and Donaldson]{RMD08}
Razick, S., Magklaras, G., and Donaldson, I.~M.
\newblock {iRefIndex}: a consolidated protein interaction database with
  provenance.
\newblock \emph{BMC Bioinformatics}, 9:\penalty0 405, 2008.

\bibitem[Turner\emph{ et~al.}(2010)Turner, Razick, Turinsky, Vlasblom, Crowdy,
  Cho, Morrison, Donaldson, and Wodak]{TRTV10}
Turner, B., Razick, S., Turinsky, A.~L. \emph{et~al.}
\newblock {iRefWeb}: interactive analysis of consolidated protein interaction
  data and their supporting evidence.
\newblock \emph{Database (Oxford)}, 2010:\penalty0 baq023, 2010.

\bibitem[Turinsky\emph{ et~al.}(2010)Turinsky, Razick, Turner, Donaldson, and
  Wodak]{TRTD10}
Turinsky, A.~L., Razick, S., Turner, B. \emph{et~al.}
\newblock Literature curation of protein interactions: measuring agreement
  across major public databases.
\newblock \emph{Database (Oxford)}, 2010:\penalty0 baq026, 2010.

\bibitem[Stojmirović and Yu(2009)]{SY09}
Stojmirović, A. and Yu, Y.-K.
\newblock {ITM Probe}: analyzing information flow in protein networks.
\newblock \emph{Bioinformatics}, 25\penalty0 (18):\penalty0 2447--9, 2009.

\bibitem[Maglott\emph{ et~al.}(2011)Maglott, Ostell, Pruitt, and
  Tatusova]{MOPT11}
Maglott, D., Ostell, J., Pruitt, K.~D. \emph{et~al.}
\newblock {Entrez Gene}: gene-centered information at {NCBI}.
\newblock \emph{Nucleic Acids Res}, 39\penalty0 (Database issue):\penalty0
  D52--7, 2011.

\bibitem[{UniProt Consortium}(2010)]{U10}
{UniProt Consortium}.
\newblock The {Universal Protein Resource} ({UniProt}) in 2010.
\newblock \emph{Nucleic Acids Res}, 38\penalty0 (Database issue):\penalty0
  D142--8, 2010.

\bibitem[Smith\emph{ et~al.}(2007)Smith, Ashburner, Rosse, Bard, Bug, Ceusters,
  Goldberg, Eilbeck, Ireland, Mungall, {OBI Consortium}, Leontis, Rocca-Serra,
  Ruttenberg, Sansone, Scheuermann, Shah, Whetzel, and Lewis]{SARB07}
Smith, B., Ashburner, M., Rosse, C. \emph{et~al.}
\newblock The {OBO Foundry}: coordinated evolution of ontologies to support
  biomedical data integration.
\newblock \emph{Nat Biotechnol}, 25\penalty0 (11):\penalty0 1251--5, 2007.

\bibitem[Tong\emph{ et~al.}(2004)Tong, Lesage, Bader, Ding, Xu, Xin, Young,
  Berriz, Brost, Chang, Chen, Cheng, Chua, Friesen, Goldberg, Haynes,
  Humphries, He, Hussein, Ke, Krogan, Li, Levinson, Lu, Ménard, Munyana,
  Parsons, Ryan, Tonikian, Roberts, Sdicu, Shapiro, Sheikh, Suter, Wong, Zhang,
  Zhu, Burd, Munro, Sander, Rine, Greenblatt, Peter, Bretscher, Bell, Roth,
  Brown, Andrews, Bussey, and Boone]{TLBD04}
Tong, A. H.~Y., Lesage, G., Bader, G.~D. \emph{et~al.}
\newblock Global mapping of the yeast genetic interaction network.
\newblock \emph{Science}, 303\penalty0 (5659):\penalty0 808--13, 2004.

\bibitem[Collins\emph{ et~al.}(2007)Collins, Kemmeren, Zhao, Greenblatt,
  Spencer, Holstege, Weissman, and Krogan]{CKZG07}
Collins, S.~R., Kemmeren, P., Zhao, X.-C. \emph{et~al.}
\newblock Toward a comprehensive atlas of the physical interactome of
  saccharomyces cerevisiae.
\newblock \emph{Mol Cell Proteomics}, 6\penalty0 (3):\penalty0 439--50, 2007.

\bibitem[Gavin\emph{ et~al.}(2006)Gavin, Aloy, Grandi, Krause, Boesche,
  Marzioch, Rau, Jensen, Bastuck, Dümpelfeld, Edelmann, Heurtier, Hoffman,
  Hoefert, Klein, Hudak, Michon, Schelder, Schirle, Remor, Rudi, Hooper, Bauer,
  Bouwmeester, Casari, Drewes, Neubauer, Rick, Kuster, Bork, Russell, and
  Superti-Furga]{GAGK06}
Gavin, A.-C., Aloy, P., Grandi, P. \emph{et~al.}
\newblock Proteome survey reveals modularity of the yeast cell machinery.
\newblock \emph{Nature}, 440\penalty0 (7084):\penalty0 631--6, 2006.

\bibitem[Krogan\emph{ et~al.}(2006)Krogan, Cagney, Yu, Zhong, Guo, Ignatchenko,
  Li, Pu, Datta, Tikuisis, Punna, Peregrín-Alvarez, Shales, Zhang, Davey,
  Robinson, Paccanaro, Bray, Sheung, Beattie, Richards, Canadien, Lalev, Mena,
  Wong, Starostine, Canete, Vlasblom, Wu, Orsi, Collins, Chandran, Haw,
  Rilstone, Gandi, Thompson, Musso, St~Onge, Ghanny, Lam, Butland, Altaf-Ul,
  Kanaya, Shilatifard, O'Shea, Weissman, Ingles, Hughes, Parkinson, Gerstein,
  Wodak, Emili, and Greenblatt]{KCYZ06}
Krogan, N.~J., Cagney, G., Yu, H. \emph{et~al.}
\newblock Global landscape of protein complexes in the yeast saccharomyces
  cerevisiae.
\newblock \emph{Nature}, 440\penalty0 (7084):\penalty0 637--43, 2006.

\bibitem[Ashburner\emph{ et~al.}(2000)Ashburner, Ball, Blake, Botstein, Butler,
  Cherry, Davis, Dolinski, Dwight, Eppig, Harris, Hill, Issel-Tarver,
  Kasarskis, Lewis, Matese, Richardson, Ringwald, Rubin, and Sherlock]{ABB00}
Ashburner, M., Ball, C.~A., Blake, J.~A. \emph{et~al.}
\newblock Gene ontology: tool for the unification of biology. {The Gene
  Ontology Consortium.}
\newblock \emph{Nat Genet}, 25:\penalty0 25--29, 2000.

\bibitem[Alves\emph{ et~al.}(2008)Alves, Ogurtsov, and Yu]{AOY08}
Alves, G., Ogurtsov, A.~Y., and Yu, Y.-K.
\newblock {RAId\_DbS}: mass-spectrometry based peptide identification web
  server with knowledge integration.
\newblock \emph{BMC Genomics}, 9:\penalty0 505, 2008.

\bibitem[Hannich\emph{ et~al.}(2005)Hannich, Lewis, Kroetz, Li, Heide, Emili,
  and Hochstrasser]{HLKL05}
Hannich, J.~T., Lewis, A., Kroetz, M.~B. \emph{et~al.}
\newblock Defining the sumo-modified proteome by multiple approaches in
  saccharomyces cerevisiae.
\newblock \emph{J Biol Chem}, 280\penalty0 (6):\penalty0 4102--10, 2005.

\bibitem[Croft\emph{ et~al.}(2011)Croft, O'Kelly, Wu, Haw, Gillespie, Matthews,
  Caudy, Garapati, Gopinath, Jassal, Jupe, Kalatskaya, Mahajan, May, Ndegwa,
  Schmidt, Shamovsky, Yung, Birney, Hermjakob, D'Eustachio, and Stein]{COWH11}
Croft, D., O'Kelly, G., Wu, G. \emph{et~al.}
\newblock Reactome: a database of reactions, pathways and biological processes.
\newblock \emph{Nucleic Acids Res}, 39\penalty0 (Database issue):\penalty0
  D691--7, 2011.

\bibitem[Blaiseau and Thomas(1998)]{BT99}
Blaiseau, P.~L. and Thomas, D.
\newblock Multiple transcriptional activation complexes tether the yeast
  activator Met4 to DNA.
\newblock \emph{EMBO J}, 17\penalty0 (21):\penalty0 6327--36, 1998.

\end{thebibliography}
%% ********************************************************************

\section*{Acknowledgments}
This work was supported by the Intramural Research Program of the National Library of Medicine at the National Institutes of Health. We thank Dr. Donaldson for his critical reading of this manuscript and for providing us with the proprietary version of iRefIndex 7.0 dataset, which was used for initial development of ppiTrim.

\newpage
\renewcommand{\tablename}{Supplementary Table} 
\setcounter{table}{0}

\begin{center}
{\Large\bf Supplementary Materials for `ppiTrim: constructing non-redundant and up-to-date interactomes'}
\end{center}
\vspace{.35cm}

\begin{center}
{\large Aleksandar Stojmirovi\'c, and Yi-Kuo Yu}
\vspace{0.25cm}
\small

\par \vskip .2in \noindent
National Center for Biotechnology Information\\
National Library of Medicine\\
National Institutes of Health\\
Bethesda, MD 20894\\
United States
\end{center}

\normalsize
\vspace{0.25cm}
\thispagestyle{plain}

\newpage

\begin{table}
\begin{center}
\caption{Description of ppiTrim MITAB 2.6 columns \label{tbl:mitabcols}}
{\footnotesize
\rowcolors{2}{gray!25}{white}
\begin{tabular}{rlp{5.0cm}>{\ttfamily\scriptsize}p{5.4cm}} \toprule
Column & Short Name & Description & {\normalfont Example} \\ \midrule
1 & uidA & Smallest Gene ID of the interactor A$^{*\dagger}$ & entrezgene/locuslink:854647 \\
2 & uidB & Smallest Gene ID of the interactor B$^*$ & entrezgene/locuslink:855136 \\
3 & altA & All gene IDs of the interactor A$^*$ & entrezgene/locuslink:854647\\
4 & altB & All gene IDs of the interactor B$^*$ &  entrezgene/locuslink:855136\\
5 & aliasA & All canonical gene symbols and integer CROGIDs of interactor A&  entrezgene/locuslink:BNR1| icrogid:2105284\\
6 & aliasB & All canonical gene symbols and integer CROGIDs of interactor B&  entrezgene/locuslink:MYO5| icrogid:3144798\\
7 & method & PSI-MI term for interaction detection method &  MI:0018(two hybrid) \\
8 & author & First author name(s) of the publication in which this interaction has been shown$^\ddagger$ &  Tong AH [2002]|tong-2002a-3\\
9 & pmids & Pubmed ID(s) of the publication in which this interaction has been shown &  pubmed:11743162\\
10 & taxA & NCBI Taxonomy identifier for interactor A &  taxid:4932(Saccharomyces cerevisiae)\\
11 & taxB & NCBI Taxonomy identifier for interactor B &  taxid:4932(Saccharomyces cerevisiae)\\
12 & interactionType & PSI-MI term for interaction type &  MI:0407(direct interaction)\\
13 & sourcedb & PSI-MI terms for source databases$^\ddagger$ & MI:0000(MPACT)|MI:0463(grid)| MI:0465(dip)|MI:0469(intact) \\
14 & interactionIdentifier & A list of interaction identifiers$^\star$ & ppiTrim:tyuGkSOK231dh3YnSi6GbczJCFE=| MPACT:8233|dip:DIP-11198E|grid:147506| intact:EBI-601565|intact:EBI-601728| irigid:288990|edgetype:X   \\
15 & confidence & A list of ppiTrim confidence scores$^\bullet$  &  maxsources:2|dmconsistency:full| conflicts:S3oaiXt5tA4vVrUsO1rc1TA9krk=\\
16 & expansion & Either `none' for binary interactions or `bipartite' for subunits of complexes &  none \\
17 & biologicalRoleA & PSI-MI term(s) for the biological role of interactor A$^\ddagger$ & MI:0499(unspecified role) \\
18 & biologicalRoleB &  PSI-MI term(s) for the biological role of interactor B $^\ddagger$ & MI:0499(unspecified role) \\
19 & experimentalRoleA & PSI-MI term(s) for the experimental role of interactor A$^\ddagger$ & MI:0496(bait)|MI:0498(prey)| MI:0499(unspecified role) \\
20 & experimentalRoleB & PSI-MI term(s) for the experimental role of interactor B$^\ddagger$ & MI:0496(bait)|MI:0498(prey)| MI:0499(unspecified role) \\
21 & interactorTypeA & PSI-MI term for the type of interactor A (either `protein' or `protein complex') & MI:0326(protein) \\
22 & interactorTypeB & PSI-MI term for the type of interactor B (always `protein') & MI:0326(protein) \\
%23 & xrefsA & Not used by ppiTrim &  - \\
%24 & xrefsB & Not used by ppiTrim  & - \\
%25 & xrefsInteraction & Not used by ppiTrim & - \\
%26 & annotationsA & Not used by ppiTrim & - \\
%27 & annotationsB & Not used by ppiTrim &  - \\
%28 & annotationsInteraction & Not used by ppiTrim & - \\
29 & hostOrganismTaxid & NCBI Taxonomy identifier for the host organism &  taxid:4932(Saccharomyces cerevisiae) \\
%30 & parametersInteraction & Not used by ppiTrim  & - \\
31 & creationDate & Date when ppiTrim was run & 2011/05/11 \\
32 & updateDate &  Date when ppiTrim was run & 2011/05/11 \\
%33 & checksumA & Not used by ppiTrim  & - \\
%34 & checksumB & Not used by ppiTrim &  - \\
35 & checksumInteraction &  ppiTrim ID for an interaction & ppiTrim:tyuGkSOK231dh3YnSi6GbczJCFE=  \\
36 & negative & Always `false' & false \\ \bottomrule
\end{tabular} 
}
\caption*{\footnotesize The above table shows short descriptions for the columns of lines output by ppiTrim with examples. The columns that are not used by ppiTrim (\texttt{-} output) are omitted. List of items are always separated by the $|$ character (without any intervening spaces). This description only applies to ppiTrim output; the full PSI-MI 2.6 TAB format description can be found at \url{http://code.google.com/p/psimi/wiki/PsimiTab26Format}{}  Notes: 
$^*$An interactor may be associated with several Gene IDs. In that case the smallest one is written in uid columns while the entire list is shown in alt columns. 
$^\dagger$Interactor A may be used to denote a protein complex. In that case the uidA is of the form \texttt{complex:$<$ppiTrim ID$>$}, while altA and aliasA are left empty. 
$^\ddagger$Multiple items are possible, originating from all source records contributing to the consolidated interaction. 
$^\star$First ID is always the ppiTrim ID for the consolidated interaction, followed by the original IDs for all contributing interactions and their integer RIGIDs from iRefIndex. The final item is the edge type code.
$^\bullet$\texttt{maxsources}: an estimate of the maximal number of independent experiments contributing to the consolidated interaction; \texttt{dmconsistency}: consistency of contributing detection method terms. Values are one of \textit{invalid} (no method terms present), \textit{single} (only one method term), \textit{min} (minimum term found but not maximum), \textit{max} (maximum term found but not minimum), and \textit{full} (both minimum and maximum term present in subcluster); \texttt{conflicts}: ppiTrim IDs of consolidated interactions with detection method term in conflict with the current one.}
\end{center}
\end{table}

\begin{table}
\begin{center}
\caption{Remapping of obsolete PSI-MI terms \label{tbl:termmappings}}
{\footnotesize
\begin{tabular}{p{0.8cm}p{5.2cm}p{0.8cm}p{5.2cm}l} \toprule
\multicolumn{2}{l}{Original Term} & \multicolumn{2}{l}{Mapped Term} & Notes \\ \midrule
MI:0021 &  colocalization by fluorescent probes cloning & MI:0428 & imaging technique & \\
MI:0022 &  colocalization by immunostaining & MI:0428 & imaging technique & $\ast$ \\
MI:0023 &  colocalization/visualisation technologies & MI:0428 & imaging technique & $\ast$ \\
MI:0025 &  copurification & MI:0401 & biochemical & \\
MI:0059 &  gst pull down & MI:0096 & pull down & \\
MI:0061 &  his pull down & MI:0096 & pull down & \\
MI:0079 &  other biochemical technologies & MI:0401 & biochemical & \\
MI:0109 &  tap tag coimmunoprecipitation & MI:0676 & tandem affinity purification & \\
MI:0045 & experimental interaction detection & MI:0492 & in vitro & $\dagger$ \\
MI:0493 & in vivo & MI:0493 & in vivo & $\dagger$ \\
MI:0000 & coip  coimmunoprecipitation & MI:0019 & coimmunoprecipitation & $\star$ \\
MI:0000 & elisa  enzyme-linked immunosorbent assay & MI:0411 & enzyme linked immunosorbent assay & $\star$ \\ \bottomrule
\end{tabular} 
}
\caption*{\footnotesize $\ast$ Interaction type is also adjusted to MI:0403 as recommended in \texttt{psi-mi.obo}; $\dagger$ HPRD terms are treated as a special case, see main text; $\star$ MPPI interactions in the human dataset.}
\end{center}
\end{table}

\begin{table}
\begin{center}
\caption{Mapping PTM labels from BioGRID into PSI-MI terms\label{tbl:ptms}}
{\footnotesize
\begin{tabular}{p{4.9cm}p{0.8cm}l} \toprule
Original Term & \multicolumn{2}{l}{Mapped Term} \\ \midrule 
Acetylation & MI:0192 & acetylation reaction \\
Deacetylation & MI:0197 & deacetylation reaction \\
Demethylation & MI:0871 & demethylation reaction \\
Dephosphorylation & MI:0203 & dephosphorylation reaction \\
Deubiquitination & MI:0204 & deubiquitination reaction \\
Glucosylation & MI:0559 & glycosylation reaction \\
Methylation & MI:0213 & methylation reaction \\
Nedd(Rub1)ylation & MI:0567 & neddylation reaction \\
No Modification & MI:0414 & enzymatic reaction \\
Phosphorylation & MI:0217 & phosphorylation reaction \\
Prenylation & MI:0211 & lipid addition \\
Proteolytic Processing & MI:0570 & protein cleavage \\
Ribosylation & MI:0557 & adp ribosylation reaction \\
Sumoylation & MI:0566 & sumoylation reaction \\
Ubiquitination & MI:0220 & ubiquitination reaction \\ \bottomrule
\end{tabular} 
}
\end{center}
\end{table}

\begin{table}
\begin{center}
\caption{Processing source interactions (RIGIDs)\label{tbl:rigids}}
{\footnotesize
\begin{tabular}{lrrrr} \toprule
Species & Initial & Without Gene ID & Retained & With Mapped Gene ID \\ \midrule
\textit{S. cerevisiae} & 186530 & 1272 & 79931 & 591 \\ 
\textit{H. sapiens} & 138570 & 1917 & 84860 & 7158 \\
\textit{D. melanogaster}  & 46925 & 4988 & 39200 & 2176 \\  \bottomrule
\end{tabular}
}
\caption*{\footnotesize Statistics of initial processing of raw interactions from in terms of iRefIndex RIGIDs. A RIGID for an interaction is a unique hash derived from its interactants' sequences (with order not significant). Thus, multiple interactions with the same interactants share the same RIGID. Shown are the initial number, number removed due to missing Gene ID, total number of retained and the number retained containing at least one interactant with mapped Gene ID. Compared to Table 2 in the main text, this table does not contain a column showing the number of removed RIGIDs due to filtering criteria. This is becuase the ppiTrim filtering routine operates on raw interactions (corresponding to a single record from a source database) and some RIGIDs would be associated with both accepted and removed raw interactions.}
\end{center}
\end{table}

\begin{table}
\begin{center}
\caption{Mapping CROGID identifiers from iRefIndex into Gene IDs: details\label{tbl:mappingdetails}}
{\footnotesize
\begin{tabular}{lrrrrrrrrrr} \toprule
Species & I & V & O & R & P & T & M & G & S & B \\ \midrule
\textit{S. cerevisiae} & 5552 & 0 & 0 & 607 & 95 & 461 & 0 & 26 & 21 & 386 \\ 
\textit{H. sapiens} & 11428 & 11 & 0 & 2615 & 155 & 2017 & 71 & 754 & 429 & 0 \\
\textit{D. melanogaster} & 7780 & 0 & 30 & 1569 & 18 & 814 & 2 & 124 & 440 & 0 \\  \bottomrule
\end{tabular}
}
\caption*{\footnotesize Detailed statistics of mapping CROGIDs into Gene IDs. All numbers denote CROGIDs: directly mapped to valid Gene IDs in the iRefIndex file (I); directly mapped to Gene IDs but the Gene IDs were updated during validation (V); directly mapped to obsolete Gene IDs (O); not directly mapped to Gene IDs -- total orphans (R); orphans with PDB accession as a primary ID (P); orphans with Uniprot accession as a primary ID (T); additionally mapped to a valid Gene ID using mapping.txt file from iRefIndex (M); additionally mapped to a valid Gene ID using a direct reference from Uniprot record (G); additionally mapped to a valid Gene ID using a gene name from Uniprot record (S); additionally mapped to a Gene ID that was not valid (B).}
\end{center}
\end{table}

\begin{table}
\begin{center}
\caption{Randomly sampled deflated complexes from high throughput publications\label{tbl:smplcmpx1}}
{\scriptsize
\rowcolors{2}{gray!25}{white}
\begin{tabular}{>{\ttfamily}lcrm{2.6cm}m{4.5cm}} \toprule
{\normalfont\scriptsize ppiTrim Complex ID} & Sources & Pubmed ID & Members & Comments \\ \midrule

8AVRUHG76vkiFn2cZGICNZzr00Y= & grid & 14759368 & CFT2, YSH1, PTA1, MPE1 & Part of mRNA cleavage/polyadenylation complex (4/10 proteins). \\
9yS57j/gbRbOlNmmimsVeonoraA= & grid & 14759368 & NUT1, MED7, MED4, SIN4, SRB4 & Part of mediator complex. \\
JU+EOkq6ipLh9DJKRtGRLUvT7vM= & grid,mint & 14759368 & UBP6, RPT3, RPN9, RPT1, RPN8, RPN2, RPN7, RPN1 & Part of proteasome. MINT does not contain complexes from the original paper. \\
HtTmhGiPyfIT2vFtRZ94uWw0rsY= & grid & 16429126 & IOC3, HTB1, HTA2, HHF2, ISW1, KAP114, ITC1, RPS4A, VPS1, NAP1, RPO31, ISW2, TBF1, BRO1, MOT1 & Part of Complex \# 99. \\
LnNzfyPGShcG7zkKynU6+fsK2eU= & grid & 16429126 & PSK1, NTH1, BMH2, RTG2, BMH1 & Part of complex \# 147 (two core proteins plus three attachments). \\
S2I6VRjFMWC6rkkM+oYXwKCg9YQ= & grid & 16429126 & RPL4B, MNN10, MNN11, HOC1, MNN9, ANP1 & Core complex (\# 111 -- mannan polymerase II) + one attachment protein (RPL4B). \\
1fRmAapl2ruoQq202YUJg55maFo= & grid,mint & 16554755 & RSM24, RSM28, MRPS5, MRP13, MRPS35, RSM27, RSM7, RSM25, MRPS17, MRPS12, RSM19, MRP4 & Part of complex \# 1. \\
5tBkYOmK/G1h3vaQmiOnUoBHHMQ= & grid,mint & 16554755 & CFT2, YSH1, MPE1, PAP1 & Part of complex \# 18. \\
9f2DVj2rDGeCP53LHOnWRMwq14A= & grid,mint & 16554755 & KAP95, RTT103, VMA2, RAI1, RAT1, RPB2, SRP1 & True experimental association but not part of any derived complex. \\
AVawv51+6Fqe3DquygD/XfyrXxE= & grid,mint & 16554755 & RRP42, RRP45, RRP6, CSL4, MPP6, RRP4, LRP1, DDI1 & Part of complex \# 19. \\
NOLEwovavMsFrQEdkSUt/mldeMc= & grid,mint & 16554755 & CDC3, SHS1, CDC11, CDC12 & Part of complex \# 121. \\
WA51i87Lj1wGp/EeF1OV/YvbW1Y= & grid,mint & 16554755 & GTT2, TRX1, CRN1, SSA3, IPP1, CMD1, TRX2, TDH1, RPL40B, CDC21, OYE2 & True experimental association but not part of any derived complex.\\
YN/hQXQvzoB5HqrgPzVth28mGsY= & grid,mint & 16554755 & RRP43, RRP42, RRP45, RRP40, DIS3, RRP6, RRP4, LRP1 & Part of complex \# 19. \\
1LRk+AgI8HpGOSAgkhDzNJWSvtI= & grid & 20489023 & RTG3, RTG2, TOR1, TOR2, CKA2, MYO2, MKS1, KOG1 & True experimental association. \\
xWzvxeJFGqjkCihjmQVf5gZhJjQ= & dip,grid,mint & 20489023 & PUF3, SAM1, GCD6, SPT16, MTC1, YGK3, LSM12 & True experimental association. \\ \bottomrule
\end{tabular} 
}
\caption*{\footnotesize To partially investigate the fidelity of deflated complexes of type A and N, we randomly sampled 25 such complexes from the final ppiTrim yeast dataset and examined the original publications associated with them. This table contains 15 deflated complexes from high-throughput publications, while Supplemenary Table~\ref{tbl:smplcmpx2} contains the complexes from low-throughput publications. Most of high-throughput papers referred to in this table present both the lists of bait-prey associations and of derived complexes. The complexes delated by ppiTrim are often derived from the former and form only parts of the latter. In the last column of this table, the complex numbers referred to are labels used by the publication's authors.}
\end{center}
\end{table}

\begin{table}
\begin{center}
\caption{Randomly sampled deflated complexes from low-throughput publications\label{tbl:smplcmpx2}}
{\scriptsize
\rowcolors{2}{gray!25}{white}
\begin{tabular}{>{\ttfamily}lcrm{2.6cm}m{4.5cm}} \toprule
{\normalfont\scriptsize ppiTrim Complex ID} & Sources & Pubmed ID & Members & Comments \\ \midrule

15VfQtoe5gxGNwPSY3AG0sq6A2U= & grid & 9891041 & CCR4, HPR1, PAF1, SRB5, GAL11 & NOT a true complex. This is because of bad annotation of PAF1--SRB5 interaction by the BioGRID. Completely opposite interpretation was given in the paper. \\
d79IdtwfTAENrH8CQ+c8CpS389Y= & grid & 10329679 & YPT1, VPS21, YPT7, GDI1 & True complex. This is the only experiment in the paper. \\
EtS4cgphEpTqJb/FS5qxyzf0ke8= & grid & 11733989 & CDC39, CCR4, CDC36, CAF130, CAF40, CAF120, POP2, NOT5, MOT2 & True complex. CAF120 is an unusual member that could almost be left out. \\
2kOyGdwzWywSpN5mhK26gCcC6LQ= & grid & 14769921 & GBP2, IMD3, TEF1, KEM1, CTK2, CTK1, CTK3 & True complex, except that TEF1 should be TEF2. This is an error in the iRefIndex source file; the BioGRID website has the correct assignment. \\
Kd07BBUF07Sqy9NP3D0lixsS/TY= & grid & 15303280 & BUD31, RPL2B, PRP19, CDC13, ATP1, RPS4A, SNU114, MDH1, MAM33, MRPL3, MRPL17, PRP8, PRP22, PAB1, BRR2 & True association \\
ZAGz/IZqkEr3/NTDLzPEDAD9cKo= & grid & 16179952 & CDC40, UFD1, SSM4, UBX2 & NOT a true complex, probably due to a typo in annotation. CDC40 cannot be found anywhere in the paper and should most likely be CDC48. \\
RDu0dsPAN0QEadfSU5sv05Ifihw= & grid & 16286007 & SIN3, RCO1, RPD3, UME1, EAF3 & True complex. \\
Vqbn3dDwTPgyE9DzbatFNqzdFe0= & grid & 16615894 & VPS36, VPS25, VPS28, SNF8 & Vps28 binds the other three, which form a complex. \\
lmdypAN9kaHBdasLWS19x8K7KkE= & grid & 20159987 & UBI4, UFD2, PEX29, SSM4 & Biological association but indicated as `NOT a stable complex' in the paper. \\
aakRh6qVahGxGvqHe399+faxPvA= & grid & 20655618 & PEX13, PEX10, PEX8, PEX12 & Association is correct, although mutant strain was used to obtain this particular complex. \\ \bottomrule
\end{tabular} 
}
\caption*{\footnotesize To partially investigate the fidelity of deflated complexes of type A and N, we randomly sampled 25 such complexes from the final ppiTrim yeast dataset and examined the original publications associated with them. This table contains 10 deflated complexes from low-throughput publications, while Supplemenary Table~\ref{tbl:smplcmpx1} contains the complexes from high-throughput publications.}
\end{center}
\end{table}

\begin{table}
\begin{center}
\caption{Summary of resolvable conflicts \label{tbl:resolvable_sce}}
{\scriptsize
\rowcolors{2}{gray!25}{white}
\begin{tabular}{>{\raggedright}p{12cm}r} \toprule
Consolidated terms & Count  \tabularnewline \midrule
MI:0018 (two hybrid), MI:0045 (experimental interaction detection), MI:0398 (two hybrid pooling approach), MI:0399 (two hybrid fragment pooling approach) & 3959 \tabularnewline 
MI:0090 (protein complementation assay), MI:0111 (dihydrofolate reductase reconstruction) & 2612 \tabularnewline 
MI:0090 (protein complementation assay), MI:0112 (ubiquitin reconstruction) & 2077 \tabularnewline 
MI:0004 (affinity chromatography technology), MI:0676 (tandem affinity purification) & 1840 \tabularnewline 
MI:0004 (affinity chromatography technology), MI:0007 (anti tag coimmunoprecipitation) & 1408 \tabularnewline 
MI:0018 (two hybrid), MI:0045 (experimental interaction detection), MI:0397 (two hybrid array) & 1231 \tabularnewline 
MI:0018 (two hybrid), MI:0045 (experimental interaction detection) & 954 \tabularnewline 
MI:0018 (two hybrid), MI:0397 (two hybrid array) & 914 \tabularnewline 
MI:0045 (experimental interaction detection), MI:0686 (unspecified method) & 628 \tabularnewline 
MI:0004 (affinity chromatography technology), MI:0019 (coimmunoprecipitation) & 598 \tabularnewline 
MI:0018 (two hybrid), MI:0398 (two hybrid pooling approach) & 506 \tabularnewline 
MI:0004 (affinity chromatography technology), MI:0007 (anti tag coimmunoprecipitation), MI:0676 (tandem affinity purification) & 444 \tabularnewline 
MI:0018 (two hybrid), MI:0045 (experimental interaction detection), MI:0686 (unspecified method) & 320 \tabularnewline 
MI:0004 (affinity chromatography technology), MI:0096 (pull down) & 217 \tabularnewline 
MI:0415 (enzymatic study), MI:0424 (protein kinase assay) & 192 \tabularnewline 
MI:0045 (experimental interaction detection), MI:0081 (peptide array) & 150 \tabularnewline 
MI:0045 (experimental interaction detection), MI:0676 (tandem affinity purification) & 120 \tabularnewline 
\midrule
MI:0492 (in vitro), MI:0493 (in vivo) & 5739 \tabularnewline 
MI:0018 (two hybrid), MI:0398 (two hybrid pooling approach) & 5394 \tabularnewline 
MI:0018 (two hybrid), MI:0492 (in vitro), MI:0493 (in vivo) & 2796 \tabularnewline 
MI:0096 (pull down), MI:0492 (in vitro), MI:0493 (in vivo) & 2760 \tabularnewline 
MI:0096 (pull down), MI:0492 (in vitro) & 2134 \tabularnewline 
MI:0018 (two hybrid), MI:0492 (in vitro) & 1658 \tabularnewline 
MI:0018 (two hybrid), MI:0493 (in vivo) & 1193 \tabularnewline 
MI:0018 (two hybrid), MI:0397 (two hybrid array) & 1045 \tabularnewline 
MI:0096 (pull down), MI:0493 (in vivo) & 513 \tabularnewline 
MI:0004 (affinity chromatography technology), MI:0006 (anti bait coimmunoprecipitation) & 384 \tabularnewline 
MI:0004 (affinity chromatography technology), MI:0019 (coimmunoprecipitation) & 309 \tabularnewline 
MI:0004 (affinity chromatography technology), MI:0007 (anti tag coimmunoprecipitation) & 195 \tabularnewline 
MI:0114 (x-ray crystallography), MI:0492 (in vitro) & 166 \tabularnewline 
MI:0004 (affinity chromatography technology), MI:0096 (pull down) & 161 \tabularnewline 
MI:0047 (far western blotting), MI:0492 (in vitro), MI:0493 (in vivo) & 106 \tabularnewline 
\midrule
MI:0018 (two hybrid), MI:0398 (two hybrid pooling approach) & 17738 \tabularnewline 
MI:0018 (two hybrid), MI:0399 (two hybrid fragment pooling approach) & 1426 \tabularnewline 
\bottomrule
\end{tabular} 
}
\caption*{\footnotesize All resolvable conflicts with counts of more than 100 for yeast (top), human (middle) and fruitfly (bottom) datasets are shown.}
\end{center}
\end{table}

\end{document}